# Energy Constraints for Climate Models from Hydrologic Partitioning

Jun Yin[1,2], Salvatore Calabrese[1,2], Edoardo Daly[3], and Amilcare Porporato[1,2]

[1]Department of Civil and Environmental Engineering, Princeton University, Princeton, New Jersey, 08544, USA.

[2]Princeton Environmental Institute, Princeton University, Princeton, New Jersey, 08544, USA.

[3]Department of Civil Engineering, Monash University, Clayton, VIC 3800, Australia.

**Abstract:**

It is well known that evaporative cooling of Earth's surface water reduces the amount of radiation that goes into sensible heat, namely the portion of radiation that produces higher temperatures. However, a rigorous use of long-term hydrologic measurements and the related theories of hydrologic partitioning have not yet been fully exploited to quantify these effects on climate. Here, we show that the Budyko's curve, a well-known and efficient framework for water balance estimation, can be effectively utilized to partition the surface energy fluxes by expressing the long-term evaporative fraction as a function of the dryness index. The combination of this energy partitioning method with hydrological observations allows us to estimate the surface energy components at watershed and continental scales. Analyzing climate model outputs through this new lens reveals energy biases due to inaccurate parameterization of hydrological and atmospheric processes, offering insight into model parameterization and providing useful information for improving climate projections.

As anticipated by Joseph Smagorinsky in his pioneering work on global atmospheric flow[1], detailed physical and biogeochemical components have gradually been incorporated into General Circulation Models (GCMs) to capture climate dynamics[2]. Increased model complexity, however, does not necessarily guarantee accurate simulation of the corresponding processes, and more comprehensive Earth System Models (ESMs) do not often lead to a reduction of the key uncertainties in climate projections[3], thus raising questions about increasing the complexity of ESMs[4]. Within this context, the partitioning of the surface energy fluxes, which is so critical to climate predictions[5], has remained one of the most vexing problems for modeling efforts[6].

The Budyko framework[7,8] is commonly adopted to partition the long-term average rainfall into runoff and evapotranspiration at catchment scales using long-term hydrological observations[9–11]. Because it involves the surface energy balance[8], this framework can be used in principle also for long-term estimates of surface-energy partitioning[12–14]. However, energy is often associated with recorded potential evapotranspiration (PET) from evaporation pans rather than net radiation, thus limiting in practice the potential use of the Budyko framework for surface-energy partitioning. As we will show, a relationship between PET and net radiation can be identified, thus permitting to overcome this problem and allowing us to connect PET to the partitioning of surface latent and sensible heat fluxes. With this additional step, new energy curves can be derived within the Budyko framework to provide robust global estimates of energy fluxes and to diagnose surface energy biases in complex ESMs.

**Hydrologic Partitioning of Surface Energy**

The Budyko curve[7,8] partitions rainfall into evapotranspiration and runoff via the dryness index $D_I = \langle \text{PET} \rangle / \langle P \rangle$ according to

$$\frac{\langle E \rangle}{\langle P \rangle} = f_B(D_I) = \left[ D_I \tanh(D_I^{-1})(1 - e^{-D_I}) \right]^{1/2}, \tag{1}$$

where $E$ is evapotranspiration, $P$ is rainfall, and the brackets refer to the long-term averages. Although other curves with extra parameters have been proposed (e.g., see review in ref. 15), it will be sufficient here to focus on the original curve proposed by Budyko. Here we show that Eq. (1) can be also used to partition latent and sensible heat flux, provided PET can be reliably linked to net radiation[12–14]. To achieve this, one has to consider that PET, as recorded by evaporation pans in long-term hydrological data, not only depends on the net radiation but is also closely related to $D_I$. As suggested by dimensional analysis (see Eq. (11) in Methods), one can reorganize Eq. (1) to express the evaporative fraction (EF) as

$$\text{EF} = \frac{f_B(D_I)}{D_I^{1-\omega}}, \tag{2}$$

where the coefficient $\omega$ is estimated to be 0.34 for the continental United States (see Methods). Eq. (2) forms our main result and hereafter is referred to as Budyko's energy curve. This curve permits the partitioning of surface latent and sensible heat fluxes from hydrological measurements of precipitation and PET. An analogous expression for the Bowen ratio (Bo) can also be obtained (see Eq. (13) in Methods).

As shown in Figure 1a and 1d, the Budyko's energy curve is bounded by the water and energy limits, which provide physical constraints for the energy partitioning. Specifically, the maximum portion of energy partitioned into latent heat flux occurs when the $D_I$ is about 0.73. In dry regions, evapotranspiration is limited by the water availability and thus EF tends to decrease with $D_I$. In wet regions, a large amount of precipitation increases the near-surface humidity and reduces the atmospheric demand for evapotranspiration. These water and energy limits roughly shape of the Budyko's energy curve such that the maximum EF is approximately located at the dry-to-wet transition regions (see dots in Figure 1d).

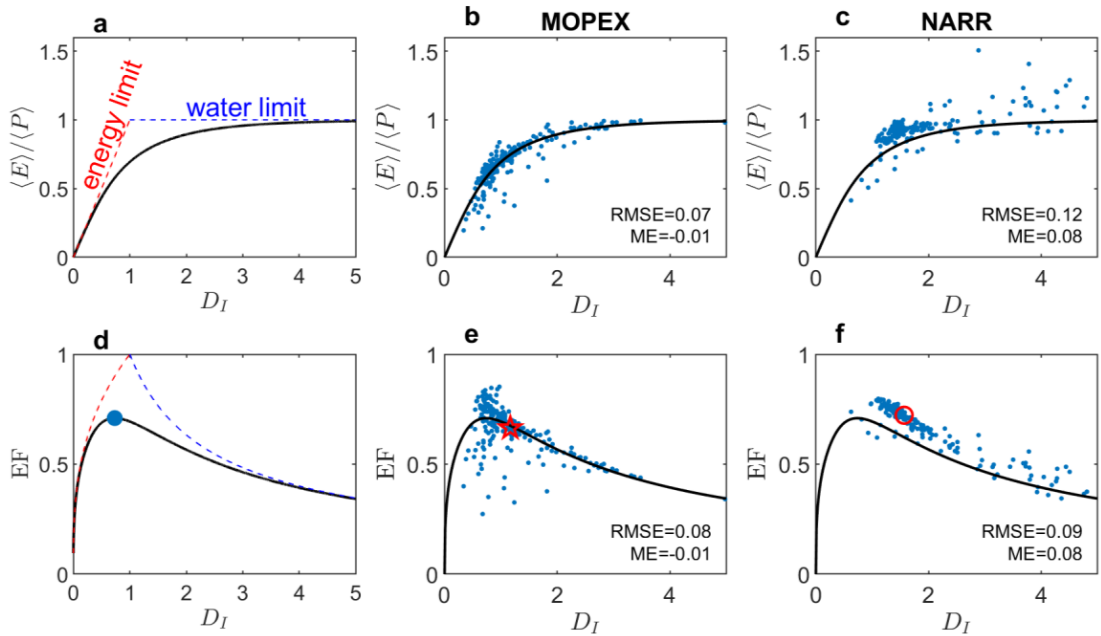

**Figure 1.** Budyko's curve for the hydrologic and energy partitioning. (a) The original Budyko curve and its application to hydrologic partitioning for (b) MOPEX and (c) NARR data. (d) The Budyko's energy curve and its application to energy partitioning for (e) MOPEX and (f) NARR data. The red and blue dash lines in panels (a) and (d) refer to the energy and water limits and the dot in (d) locates the maximum EF at the dryness index of approximately 0.73. Each point in panels (b), (c), (e), and (f) refers to one watershed from MOPEX with the size larger than 2048 km$^2$ (see thin solid boundaries in Figure 3), which is at least twice the grid size of the NARR data. The star in (e) and circle in (f) refer to the region as delineated by thick dotted lines in Figure 3 (also see the zoom-in plot in Figure 4a).

With the Budyko's energy curves so obtained, we can then partition the surface net radiation into sensible and latent heat fluxes for any given long-term dataset of precipitation and PET. Here, we use data from the international Model Parameter Estimation Experiment (MOPEX)[16] and North American Regional Reanalysis (NARR)[17]. MOPEX is designed for land surface parameterization of atmospheric and hydrological models, whereas NARR uses high-resolution model and data assimilation system to provide more accurate hydrometeorological data. For a meaningful comparison, we focus on watersheds defined in MOPEX with an area at least twice the size of NARR grids; this is consistent with the size of the watersheds used to derive the original Budyko curve[8].

As shown in Figure 1b and 1e, data from MOPEX are close to both Budyko's water and energy curves, demonstrating their usefulness in partitioning not only the water but also the energy fluxes. Particularly, the corresponding deviations from the curves tend to be smaller over the larger watersheds (see Supplementary Fig. 1) and their averages are close to zero (see statistics of mean errors, ME, in Figure 1). Given that we are interested in the large-scale energy partitioning for climate modeling, there is no need here to use more sophisticated versions of Budyko curve (e.g., ref. 10). For the NARR data, most of the corresponding points fall above the original Budyko curve (Figure 1c), causing systematic deviations from the energy curve (i.e., higher EF in Figure 1f). The statistics of root mean squared errors (RMSE) and mean errors (ME) of the deviations further suggest NARR is consistently biased from small to large scales.

To explain these biases, we compared water and energy components from NARR with those from MOPEX and Clouds and the Earth's Radiant Energy System (CERES)[18]. Evapotranspiration from MOPEX is calculated as the long-term observed precipitation subtracted by runoff, which has been regarded as the best available long-term estimate of evapotranspiration at catchment and global scales[19,20]. Radiation from CERES is often used as one of the important references for evaluating energy balance in climate model outputs[6]. The comparison of water and energy components shows that the most significant biases come from the overestimation of evapotranspiration (Figure 2 a), whereas precipitation from NARR is consistent with the MOPEX observations (Figure 2 b) and net radiation from NARR is slightly larger than that from CERES satellite data (Figure 2 c). The net radiation biases, when converted to water flux by latent heat of vaporization, are approximately 1/3 the magnitude of the evapotranspiration biases.

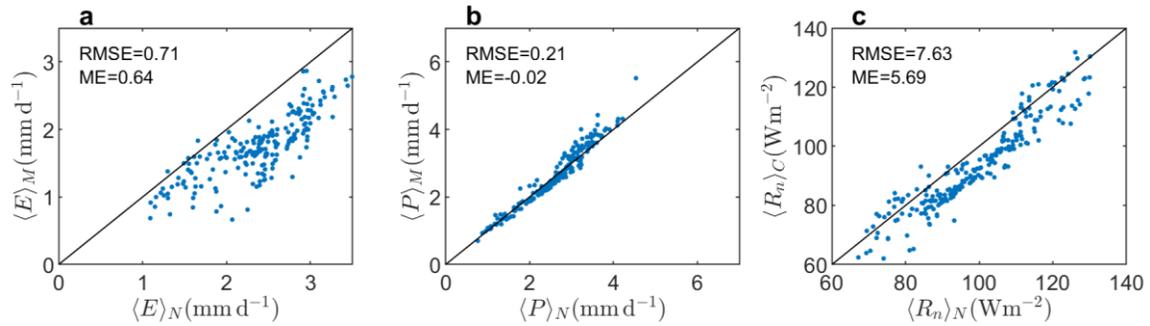

**Figure 2**. Bias analysis for water and energy balance from NARR data. Comparison of evapotranspiration (a), precipitation (b), and net radiation (c) from NARR (denoted by subscript N), MOPEX observations (denoted by subscript M), and CERES satellite data (denoted by subscript C)[18]. The solid lines are the 1:1 lines; the dash lines represent the fitted linear function with the coefficients in 95% confidence intervals and statistics shown in the corresponding panels. Each point refers to one selected watershed (as in Figure 1 b, c, e, and f).

To analyze the spatial distribution of the energy partitioning, we compared EF from NARR data with EF calculated using the Budyko's energy curve. While computed from grid data, the spatial patterns of EF in Figure 3 may be interpreted over a relatively large area to represent watershed-scale energy partitioning. Both maps show similar spatial patterns with larger EF over the east and northwest, lower values over the southwest, and strong gradients across the Great Plains. EF calculated from the Budyko's energy curve is slightly lower (lighter color in Figure 3b), consistently associated with the identified biases from the selected watersheds (see Figure 1e). These biases introduce larger latent heat fluxes, possibly intensifying atmospheric convection at continental scale[21,22].

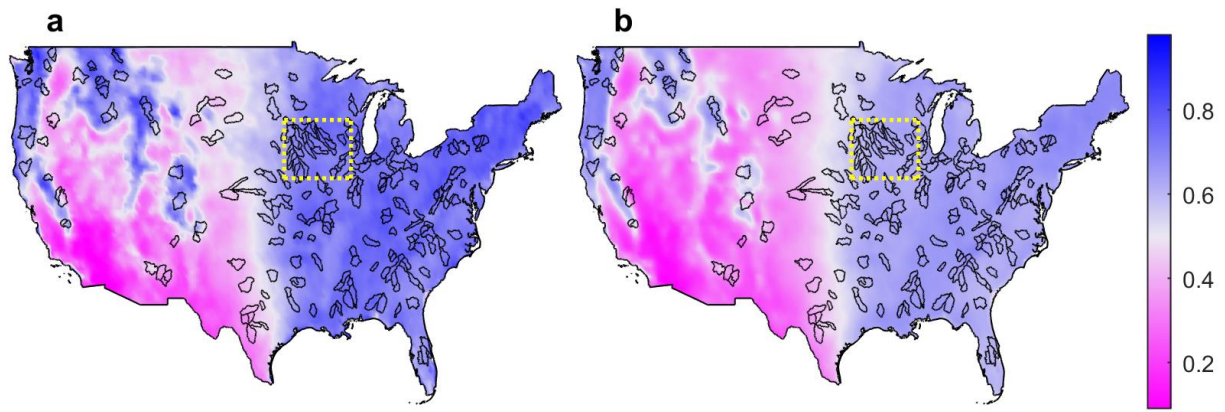

**Figure 3**. Geographic distribution of evaporative fraction from reanalysis data. Evaporative fraction is (a) directly calculated from NARR data and (b) estimated from the Budyko's energy curve with dryness index from NARR data. The black lines delineate the boundaries of the studied watersheds of MOPEX with the size larger than 2048 km$^2$ (at least twice the size of NARR grid sizes). The yellow dotted rectangle delineates the boundaries of the studied region for biases analysis (see Figure 4 and the related text).

**Energy Constraints for Climate Models**

The persistent biases at both watershed and continental scales observed in NARR data (see Figure 1-3) raise questions about their origin as well as whether climate models may also experience analogous biases. To answer this question, we treated the Budyko's energy curve and hydrological observations (here, MOPEX data are used) as correct references. Datasets that deviate from the Budyko curve suggest misrepresentation of hydrological processes possibly caused by biases in land-surface parameterization (e.g. $\Delta EF$ as denoted by the vertical dashed lines in Figure 4a). On the other hand, any deviation from the observed $D_I$ suggest biases in precipitation and/or PET likely caused by inaccurate modeling of atmospheric circulation (e.g. $\Delta D_I$ as denoted by the horizontal dashed lines in Figure 4a). These two metrics help identifying the origins of model biases as explained in the following detailed examples.

We focused on the large area over Iowa and Wisconsin delimited by the dotted line in Figure 3, where there is a dense hydrological observation network. Given that the energy partitioning in this region is relatively homogenous (see Figure 3), we average the MOPEX observations within the region to calculate the corresponding $D_I$ and EF (see the star in Figure 1 and Figure 4). These MOPEX averages are then compared with climate model outputs and reanalysis data within the Budyko's energy framework. Figure 4 reports $D_I$, EF, the biases in the estimation of the dryness index ($\Delta D_I$), and the deviations from the Budyko curve ($\Delta EF$). As can be seen, INM-CM4 is very close to the MOPEX observations in terms of both $D_I$ and EF. Data from BNU-ESM, INM-CM4, IPSL-CM5A, CSIRO-Mk3.6.0, and HadGEM2-ES are close to the Budyko's energy curve. These models may well describe the surface hydrological processes but the last three are away from the observed dryness index likely because of atmospheric circulation biases. Data from GISS-E2-H, MRI-CGCM3, ACCESS-1.3, NorESM1-M, and NARR deviate from the Budyko's energy curve, indicating their land surface parameterization may require adjustments. In terms of energy partitioning, some climate models shows large biases with $\Delta EF$ in the order of 0.1, which may determine the transition from shallow to deep atmospheric convection[21–23] and is close to the magnitude of the terrestrial energy uncertainties in climate models[6].

These results show that only a few ESMs capture the long-term water and energy partitioning embedded within the Budyko's curves and present in the MOPEX data. As the new generation climate models started to account for more climate components and sub-grid physical processes, their corresponding parameterization becomes more difficult. The long-term hydrological records coupled to the Budyko's energy curves can be used to identify the origin of the biases and provide a better representation of the mean states of the climate system. This large spatial scale averaging imposes robust physical constraints on ESMs and can be possibly used to guide both stochastic and deterministic parameterization for climate modeling[24,25].

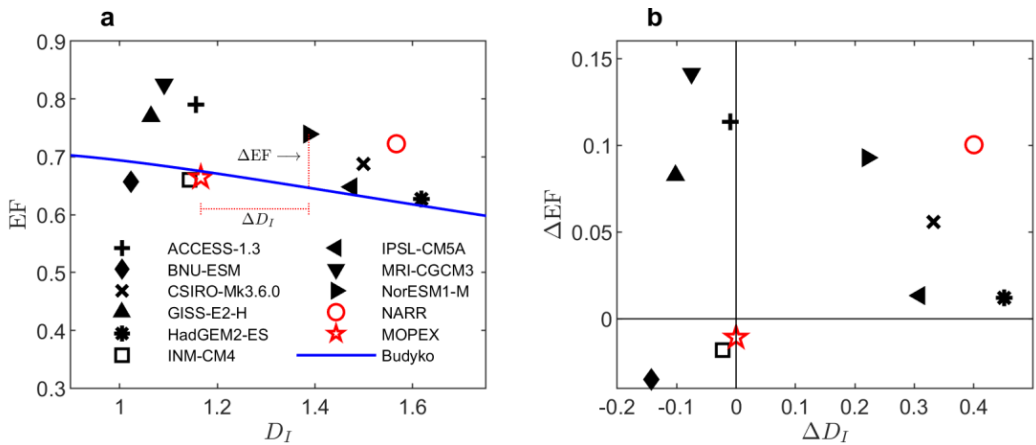

**Figure 4**. Diagnosing energy partitioning biases in climate models using the Budyko's energy curve. Panel (a) shows the relationship between evaporative fraction (EF) and dryness index ($D_I$) calculated from different datasets over Iowa and Wisconsin in the United States (see the dotted lines in Figure 3). The solid blue line is the Budyko's energy curve. The deviations of EF from the Budyko curve (shown as thin dash line and denoted as $\Delta$EF) and the deviations of dryness index from MOPEX observations (denoted as $\Delta D_I$) are presented in panel (b).

In summary, by establishing a relationship between potential evapotranspiration and net radiation, we derived the Budyko's energy curves, expressing the evaporative fraction as a function of the dryness index. Within this framework, we analyzed the geographic features of surface energy partitioning over the continental United States, showing that NARR overestimates surface latent heat fluxes. Application to climate model outputs illustrates the energy partitioning biases that may result from imperfect parameterization of land surface and atmospheric models. The identified origins of biases allow us to focus on improving the specific components of the ESMs represented by a tradeoff between model complexity and difficulty in the prediction of long-term hydroclimatic fluxes.

**Acknowledgments:**

We acknowledge support from the USDA Agricultural Research Service cooperative agreement 58-6408-3-027; and National Science Foundation (NSF) grants EAR-1331846, EAR-1316258, FESD EAR-1338694 and the Duke WISeNet Grant DGE-1068871.

**Author contributions**

J.Y., S.C., E.D. and A.P. conceived and designed the study. J.Y. wrote an initial draft of the paper, to which all authors contributed edits throughout.

**Competing interests**

Authors declare no competing interests.

**Methods**

**Budyko's Energy Curve**

Potential evapotranspiration (PET) can be modeled using Penman equation [26]

$$\rho_w \lambda_w \text{PET} = \frac{\Delta R_n}{\gamma + \Delta} + \frac{1}{\gamma + \Delta} \rho c_p g_a \text{VPD}, \tag{3}$$

where $\rho_w$ is water density, $\rho$ is air density, $c_p$ is the specific heat capacity of air, $\lambda_w$ is the latent heat of vaporization, $R_n$ is net radiation, $g_a$ is aerodynamic conductance, $\gamma$ is psychrometric constant, VPD is vapor pressure deficit, defined as

$$\text{VPD} = (1 - \text{RH})e_{\text{sat}}(T), \tag{4}$$

where RH is relative humidity and $e_{\text{sat}}$ is saturated water vapor pressure expressed as a function of air temperature, $T$, and $\Delta$ is the slope of saturation vapor pressure curve, which can be calculated from the Clausius-Clapeyron equation

$$\Delta = \frac{\lambda e_{\text{sat}}(T)}{R_w T^2}, \tag{5}$$

where $R_w$ is the specific gas constant of water. Penman equation (3) accounts for both the equilibrium evaporation under well-watered condition and evaporation due to vapor pressure deficit. This definition of PET is consistent with pan evaporation as often recorded in long-term hydrological data sets[14,27,28]. Substituting (4) and (5) into (3) yields another form of Penman equation

$$\rho_w \lambda_w \text{PET} = \frac{\Delta}{\gamma + \Delta}\left[R_n + \frac{R_w T^2}{\lambda_w}\rho c_p g_a (1\text{-RH})\right]. \tag{6}$$

Being interested in long-term averages (indicated with brackets), we can approximate the Penman equation as

$$\rho_w \lambda_w \langle \text{PET} \rangle \approx \frac{\langle \Delta \rangle}{\gamma + \langle \Delta \rangle} \left[ \langle R_n \rangle + \underbrace{\frac{R_w \langle T \rangle^2}{\lambda_w} \rho c_p g_a \left(1 - \langle \text{RH} \rangle \right)}_{\phi} \right] = \rho_w \lambda_w \, \text{PET} \,, \quad (7)$$

where $\phi$ has unit of flux (Wm$^{-2}$) and the aerodynamic conductance is assumed to be constant ($g_a$ = 18 mm s$^{-1}$). This is a reasonable approximation as shown in Supplementary Fig. 2a, which compares $\langle \text{PET} \rangle$ (i.e., the long-term average) and $\overline{\text{PET}}$ (i.e., Eq. (7)) over the continental United States from NARR datasets (32×32 km grids).

The relative humidity, $\langle \text{RH} \rangle$, appearing in Eq. (7) is often used to estimate precipitation [29–31]

$$\langle P \rangle = a \langle \text{RH} \rangle^b \,, \quad (8)$$

where $a$ and $b$ are empirical coefficients. This relationship is shown in Supplementary Fig. 2b with fitted power-law function. Combining (7), (8), and the definition of dryness index ($D_I = \langle \text{PET} \rangle / \langle P \rangle$) yields an implicit function

$$f_1 \left( \rho_w \lambda_w \langle \text{PET} \rangle, \langle R_n \rangle, \phi, D_I \right) = 0 \,, \quad (9)$$

where the first three variables have dimensions of heat fluxes while the last one is dimensionless. Applying dimensional analysis[32] yields

$$\beta = \frac{\rho_w \lambda_w \langle \text{PET} \rangle}{\langle R_n \rangle} = f_2 \left( \frac{\phi}{\langle R_n \rangle}, D_I \right) \,, \quad (10)$$

which suggests that the ratio of long-term average PET to $R_n$ can be modeled as a function of $\phi / \langle R_n \rangle$ and $D_I$. Since temperature and radiation are highly correlated, $\phi / \langle R_n \rangle$ cancels most of its information such that it has limited contribution to the variation of $\beta$ (see Supplementary

Fig. 2c). This allows us to model $\beta$ only using $D_I$, for example with power-law functions as illustrated in Supplementary Fig. 2d. As shown in the red curve, the scaling factor is very close to unity, and, for simplicity, we use

$$\beta = \frac{\rho_w \lambda_w \langle \text{PET} \rangle}{\langle R_n \rangle} = D_I^\omega, \tag{11}$$

where the coefficient $\omega$ is estimated to be 0.34. To verify this relationship, we re-plot it by combining hydrological data from MOPEX or from Climatic Research Unit (CRU) with net radiation data from CERES, showing consistent patterns as that from NARR datasets (see Supplementary Fig. 3a). The re-estimated coefficient $\omega$ has certain variations but does not have qualitative impacts on our analysis of surface energy partitioning in climate models (see Supplementary Fig. 3b).

We are now ready to derive the Budyko's energy curve by expressing evaporative fraction (EF) as the ratio of the long-term average latent heat flux to the net radiation,

$$\text{EF} = \frac{\rho_w \lambda_w \langle E \rangle}{\langle R_n \rangle} = \frac{\rho_w \lambda_w \langle E \rangle / \langle P \rangle}{\langle R_n \rangle / \langle P \rangle} = \frac{\frac{\rho_w \lambda_w \langle \text{PET} \rangle}{\langle R_n \rangle} \langle E \rangle / \langle P \rangle}{\langle \text{PET} \rangle / \langle P \rangle}. \tag{12}$$

Combining (11), (12), definition of dryness index, $D_I = \langle \text{PET} \rangle / \langle P \rangle$, and the Budyko curve, $\langle E \rangle / \langle P \rangle = f_B(D_I)$ yields Eq. (2) in the main text. This equation describes the Budyko's energy curve and have been used in the main text for partitioning surface energy fluxes. Analogously, one can express the Bowen ratio as

$$\text{Bo} = \frac{1}{\text{EF}} - 1 = \frac{D_I^{1-\omega}}{f_B(D_I)} - 1, \tag{13}$$

which also can be used for partitioning surface energy (see Supplementary Fig. 4).

**Statistical Analysis**

To evaluate the accuracy of the data, we use the metrics mean error (ME) and root mean squared error (RMSE). The errors are calculated as the deviation from the Budyko's curve (see Figure 1) or as the differences between the NARR data and the corresponding reference data (see Figure 2). The mean of the errors is referred to as ME and the uncorrected sample standard deviation of the errors is RMSE. ME is used to evaluate the overall bias of the data and RMSE is used to measure the average deviation of each individual data point. To find the best-fitting curve, we estimate the coefficients of the functions by minimizing the sum of the squares of the residuals of the points from the curve. Statistics, such as coefficient of determination ($R^2$) and coefficients in 95% confidence intervals, are presented to quantify the goodness of fit of the corresponding model.

**Data Availability**

The MOPEX data are available from NOAA's National Weather Service (www.nws.noaa.gov/ohd/mopex/mo_datasets.htm); the NARR data can be obtained from NOAA Earth System Research Laboratory (www.esrl.noaa.gov/psd/data/gridded/data.narr.html); the climate model data can be downloaded from the fifth phase of the Coupled Model Intercomparison Project website (http://cmip-pcmdi.llnl.gov); the Climatic Research Unit (CRU) TS v. 4.02 data are available at http://www.cru.uea.ac.uk/data/; the Clouds and the Earth's Radiant Energy System (CERES) EBAF-Surface can be ordered from https://ceres.larc.nasa.gov/.

Supplementary Information for

# Energy Constraints for Climate Models from Hydrologic Partitioning

Jun Yin, Salvatore Calabrese, Edoardo Daly, and Amilcare Porporato

The following figures provide complementary information regarding the Budyko's energy curve and supporting data for the section of Methods

- Supplementary Fig. 1 shows how the size of watershed influences water and energy partitioning in Budyko's curves.
- Supplementary Fig. 2 uses NARR data to support the dimensional analysis in the Method section.
- Supplementary Fig. 3 uses extra data to support the methods.
- Supplementary Fig. 4 shows extra results for surface energy partitioning.

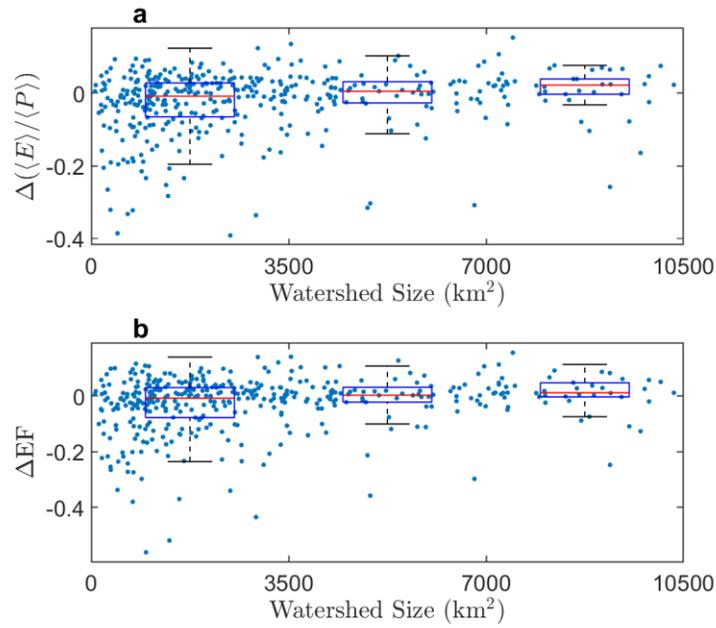

**Supplementary Fig. 1**. Relationships between the size of the watersheds and the corresponding (a) deviations from the original Budyko's curve (i.e., the errors in Figure 1b) and (b) deviations from the Budyko's energy curve (i.e., the errors in Figure 1e). The three boxplots in each panel show the statistics (25th, 50th, 75th percentiles, and extremes) of the deviations within the watersheds of the size 0-3500, 3500-7000, and 7000-10500 km², respectively.

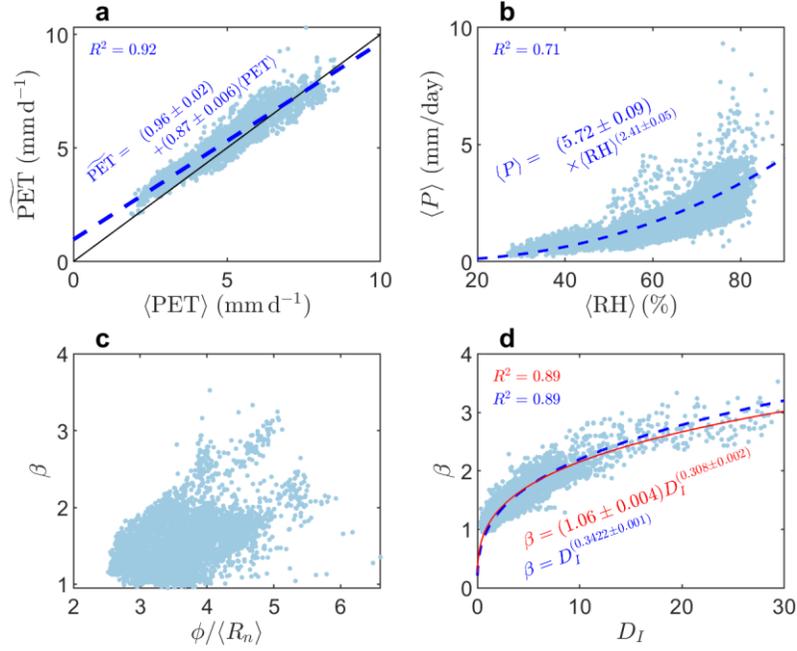

**Supplementary Fig. 2**. Long-term meteorological data from NARR over the continental United States. (a) Comparison between long-term average PET and PET calculated from Eq. (7). (b) The relationship between long-term average precipitation and near-surface relative humidity. (c) The relationship between $\beta$ and $\phi/\langle R_n \rangle$ (see Eq. (10)). (d) The relationship between $\beta$ and $D_I$. The blue dash lines and red solid line represent the fitted functions with the coefficients in 95% confidence intervals and statistics shown in the corresponding panels. Note that the 1:1 line is shown in (a) as the solid black line and two fitted power-law functions are given in (d).

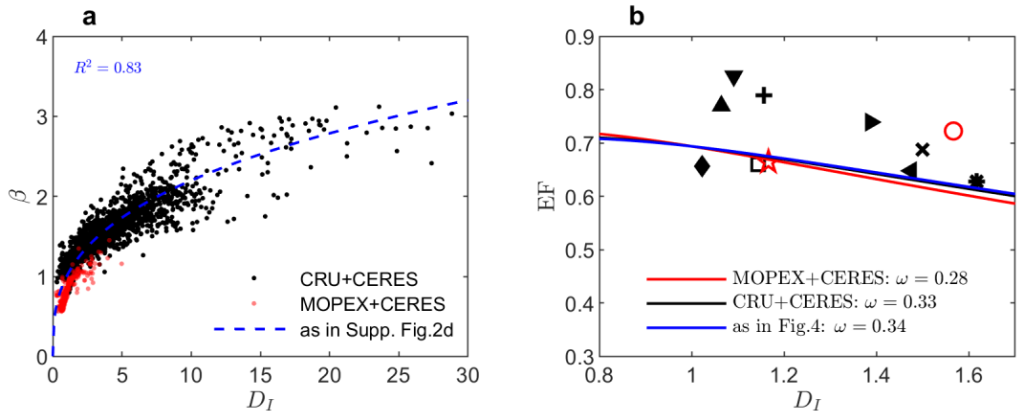

**Supplementary Fig. 3**. (a) As in Supplementary Fig. 2d but for different datasets. (b) As in Figure 4a but with $\omega$ calibrated from different datasets. Hydrological data from MOPEX or CRU are combined with radiation data from CERES to calibrate the parameter $\omega$, which is then substituted into Eq. (2) to find the corresponding Budyko's energy curve as presented in (b). To be consistent with the records from evaporation pan, Penman equation of Eq. (3) is used to calculated PET with other meteorological variables from CRU.

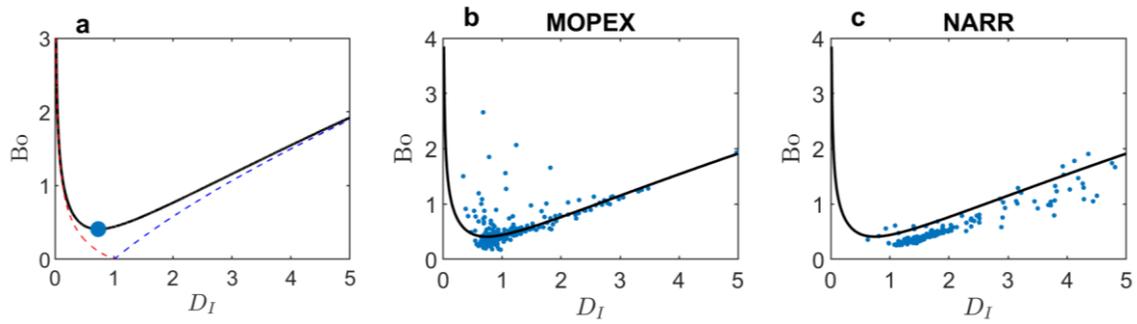

**Supplementary Fig. 4**. As in Figure 1 in the main text but for Bowen ratio.